\documentclass{PoS}
\usepackage{epsfig}
\newcommand{\be}{\begin{equation}}
\newcommand{\ee}{\end{equation}}
\def\slashed{{/}\mskip-11.0mu}
\PoS{PoS(LAT2005)229}

\title{Perturbative renormalization in parton distribution functions 
using improved actions}

\ShortTitle{Perturbative renormalization in parton distribution 
functions }

\author{\speaker{Mihalis Ioannou}\\
        Univ of Cyprus; Physics Dept\\
        E-mail: \email{mike80bi@yahoo.gr}}

\author{Haralambos Panagopoulos\\
        Univ of Cyprus; Physics Dept\\
        E-mail: \email{haris@ucy.ac.cy}}

\abstract{We calculate the 1-loop renormalization of a set of extended
  fermionic bilinears which form a basis corresponding to moments of
  the parton distribution functions. 

 We use the overlap action for fermions and Luescher-Weisz (LW) action
 for gluons. 

 Our results are presented as a function of the overlap parameter
 $\rho$ and the parameters entering the LW  action.}

\FullConference{XXIIIrd International Symposium on Lattice Field 
Theory\\
		 25-30 July 2005\\
		 Trinity College, Dublin, Ireland}

\begin{document}

\section{Introduction.}
 The description of physics in the Bjorken limit involves the operator product expansion which has the standard form:
\be
J(x)J(0)\sim\sum_{n,i,l}C_l^{n,i}(x^2)x^{\mu_1}...x^{\mu_n}O_{\mu_1...\mu_n}^{(n,i)}(0)
\ee

The forward matrix elements of the local operators $O^{(n,i)}$ appearing in this expansion are directly related to the moments of hadron structure functions.
   The dominant contribution in the expansion is given by operators whose twist (dimension minus spin) equals two, which in the flavor non-singlet case means the symmetric traceless operators\cite{ref1}
\be
O_{\mu\mu_1...\mu_n}=\bar{\psi}\gamma_{\{\mu} D_{\mu_1}...D_{\mu_n\}}\frac{\lambda^a}{2}\psi
\ee
\be
O_{\mu\mu_1...\mu_n}^{(5)}=\bar{\psi}\gamma_{\{\mu}\gamma_5 D_{\mu_1}...D_{\mu_n\}}\frac{\lambda^a}{2}\psi
\ee
where $\lambda^a$ are flavor matrices.
In this work we compute the renormalization factors of the following quantities on the lattice.
\begin{itemize}
\item[a)] Self-energy.
\be
S_N^{-1}=i\gamma_\mu P_\mu (1-\frac{g^2C_F}{16\pi^2}\Sigma_1)
\ee
\item[b)] Quark bilinears.
\be
O_X=\bar{\psi}(x)\Gamma^X\psi(x)
\ee
\item[c)] Operators which measure the first moment of quark momentum distributions.
\be
O_{V_2}=\bar{\psi}\gamma_{\{1}D_{4\}}\psi
\ee
\be
{O'}_{V_2}=\bar{\psi}\gamma_4 D_4 \psi-\frac{1}{3}\sum_{i=1}^3\bar{\psi}\gamma_i D_i\psi
\ee
\item[d)] Operators which measure the first moment of quark helicity distributions.
\be
O_{A_2}=\bar{\psi}\gamma_{\{1}\gamma_5D_{4\}}\psi
\ee
\be
{O'}_{A_2}=\bar{\psi}\gamma_4\gamma_5 D_4 \psi-\frac{1}{3}\sum_{i=1}^3\bar{\psi}\gamma_i\gamma_5 D_i\psi
\ee
\end{itemize}
Work is in progress using operators which measure the second moment of quark distributions.
A longer write-up of our results can be found in our Ref.\cite{ref2}.
\section{Calculational details.}
We denote the lattice action by
\be
S=S_G+S_F
\ee
where $S_G$  is the gluon action, and $S_F$  is the fermion action.
The gluon action we consider is written in standard notation:
\be
S_{gl}=-\frac{1}{g^2}\bigl(c_0\sum_{plaquette}TrU_{pl}+c_1\sum_{rectangle}TrU_{rtg}+c_2\sum_{chair}TrU_{chr}+c_3\sum_{parallelogram}TrU_{plg}\bigr)
\ee
where $U_{pl}$  is the standard plaquette, while the remaining U's cover all possible closed loops containing up to six links.
The coefficients $c_0$, $c_1$, $c_2$, $c_3$  satisfy the normalization condition:
\be
c_0+8c_1+16c_2+8c_3=1
\ee
As in Ref.\cite{ref3}, we have used the values of $c_0$, $c_1$, $c_2$, $c_3$, shown in Table I, where ``Plaquette'' is the standard Wilson action for gluons, ``Symanzik'' is the  tree-level improved action \cite{ref4} and TILW is the  tadpole improved Luescher-Weisz action \cite{ref5,ref6}.
\begin{table}[ht]
\begin{center}
\label{values of coefficients}
\begin{tabular}{|c|r@{}l|r@{}l|r@{}l|r@{}l|}
\multicolumn{1}{c}{action}&
\multicolumn{2}{c}{$c_0$} &
\multicolumn{2}{c}{$c_1$} &
\multicolumn{2}{c}{$c_2$} &
\multicolumn{2}{c}{$c_3$} \\
\hline
Plaquette  &1&.0   &0&.0	        &0&.0	     &0&.0 \\
Symanzik   &1&.66666	&-0&.083333	&0&.0	     &0&.0 \\
TILW1 $\beta=8.60$         &2&.31681     &-0&.151791		&0&.0	     &-0&0128098 \\
TILW2 $\beta=8.45$       &2&.34602         &-0&.154846	&0&.0	     &-0&.0134070 \\
TILW3 $\beta=8.30$        &2&.38698	        &-0&.159128		&0&.0	     &-0&.0142442 \\
TILW4 $\beta=8.20$       &2&.41278	        &-0&.161827	&0&.0	     &-0&.0147710 \\
TILW5 $\beta=8.10$        &2&.44654   &-0&.165353	&0&.0  &-0&.0154645 \\
TILW6 $\beta=8.00$       &2&.48917   &-0&.169805	&0&.0 &-0&.0163414 \\
Iwasaki [7]      &3&.648		&-0&.331	&0&.0	     &0&.0 \\
DBW2 [8]       &12&.2688     	&-1&.4086	&0&.0  &0&.0 \\
\hline
\end{tabular}
\end{center}
\caption{The values of coefficients $c_0, c_1, c_2, c_3$}
\end{table}

The action for massless overlap fermions is given by \cite{ref9}
\be
S_f=\sum_{f}\sum_{x,y}\bar{\psi}_x^f D_N(x,y)\psi_y^f
\ee
with
\be
D_N=\rho[1+X(X^{\dag}X)^{-\frac{1}{2}}]
\ee
and: $X=D_w-\rho$.  Here $D_w$  is the massless Wilson-Dirac operator with $r = 1$, and $\rho$ is a free parameter whose value must be in the range $0<\rho<2$  in order to guarantee the correct pole structure of $D_N$.
 \section{Calculations and Results.}
\noindent
{\bf Self energy}

Let us consider the massless quark propagator $S_N$  first. The inverse of $S_N$  can be written
\be
S_N^{-1}=i\gamma_{\mu}p_{\mu}(1-\frac{g^2C_F}{16\pi^2}\Sigma_1)
\ee
with $C_F=(N^2-1)/2N$, and $\Sigma_1(a,p)=log(a^2p^2)+b_{\Sigma}$ (Feynman gauge). Our results for $b_{\Sigma}$ are shown in Fig.1.\\

\noindent
{\bf Quark bilinears}

Let us consider local operators of the form
\be
O_X=\bar{\psi}(x)\gamma^X\psi(x)
\ee
with $\Gamma^S=1,\ \Gamma^P=\gamma_5,\ \Gamma^V=\gamma_{\mu},\ \Gamma^A=\gamma_{\mu}\gamma_5,\ \Gamma^T=\sigma_{\mu\nu}\gamma_5$ ,
i.e. $X=S,\ P,\ V,\ A$ and $ T.$
We denote the amputated Green's function of the operator $O_X$ by $\Lambda^X.$ 
The final results are (in Feynman gauge):
\be
\Lambda^{S,P}=\{1,\gamma_5\}+\frac{g^2C_F}{16\pi^2}[-4log(a^2p^2)+b_{S,P}]\{1,\gamma_5\}
\ee
\be
\Lambda_{\mu}^{V,A}=\{\gamma_\mu,\gamma_\mu\gamma_5\}+\frac{g^2C_F}{16\pi^2}[-\gamma_\mu(log(a^2p^2)+b_{V,A})+2\frac{p_\mu \slashed {p}}{p^2}]\{1,\gamma_5\}
\ee
\be
\Lambda_{\mu,\nu}^T=\sigma_{\mu,\nu}\gamma_5+\frac{g^2C_F}{16\pi^2}b_T\sigma_{\mu,\nu}\gamma_5
\ee
Where $b_S=b_P$ and $ b_V=b_A$.
Only one diagram (``Vertex'') contributes to $\Lambda^X$.

Our results for $b_{\Sigma},\  b_{S,P},\  b_{V,A}$  and  $ b_T $ provide a cross check and an extension of results appearing in Refs.\cite{ref10,ref3}. They are shown in Figures 1, 2, 3, 4 for typical values of the overlap parameter $\rho$.

\begin{figure}
\hspace{-1.5cm}
\begin{minipage}[t]{0.45\linewidth}
\begin{center}
\epsfig{file=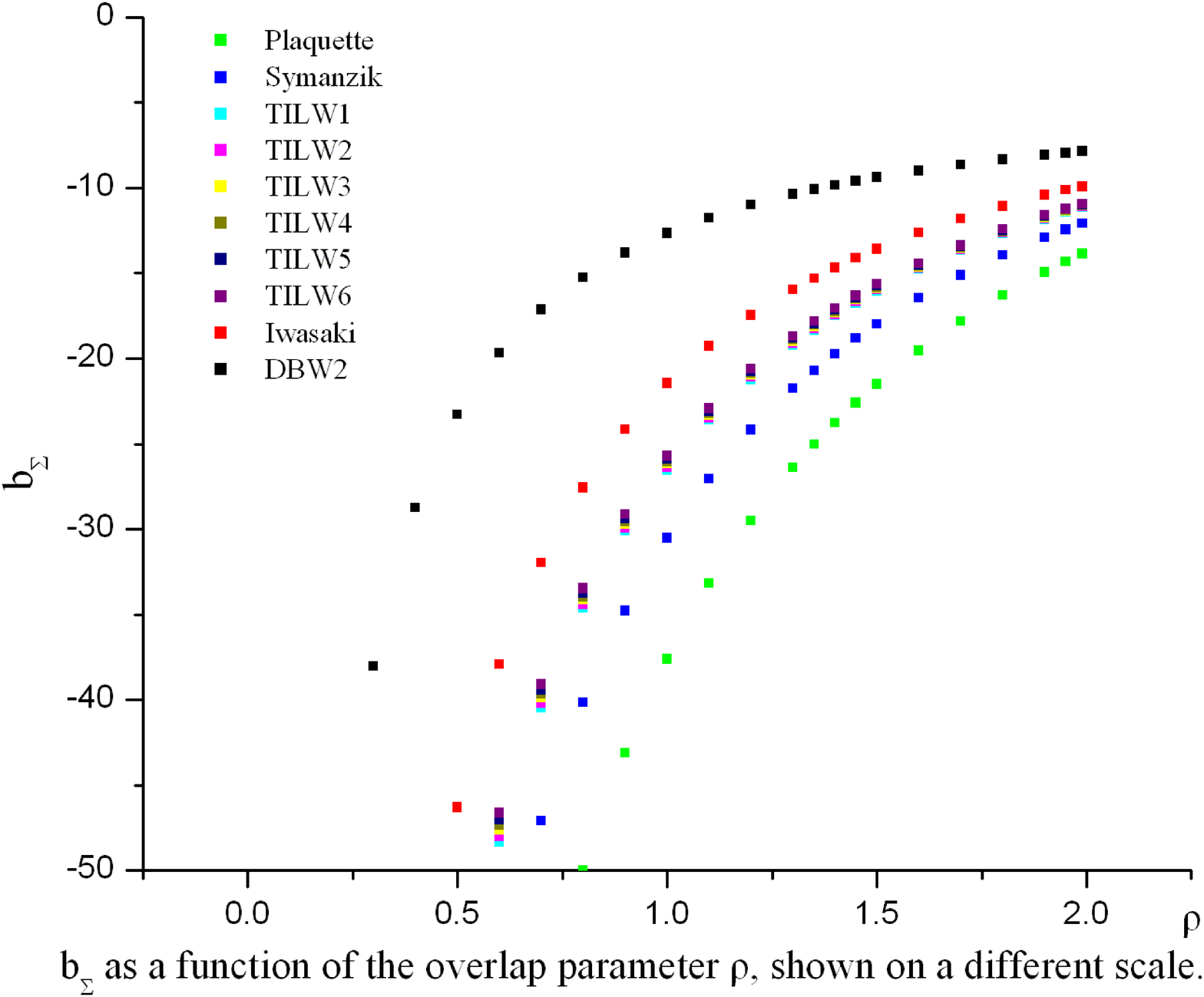, scale=0.15}
\label{selfpicnew}
\hfill\\
\hspace{2cm}{\small\bf Figure 1}
\end{center}
\end{minipage}
\hspace{1.3cm}
\begin{minipage}[t]{0.45\linewidth}
\begin{center}
\epsfig{file=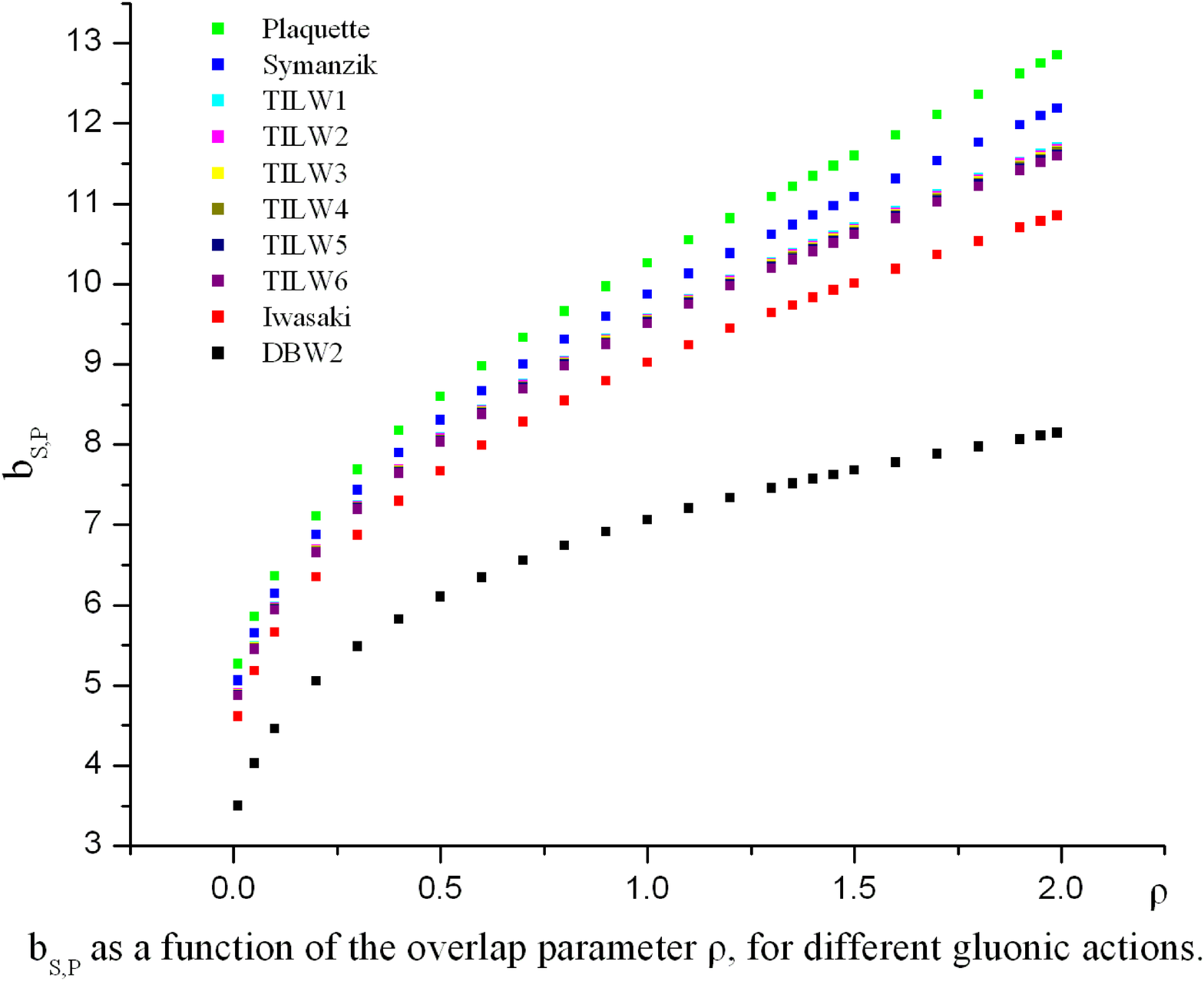, scale=0.15}
\label{scalarpic}
\hfill\\
\hspace{2cm}{\small\bf Figure 2}
\end{center}
\end{minipage}
\end{figure}
\begin{figure}
\hspace{-1.5cm}
\begin{minipage}[t]{0.45\linewidth}
\begin{center}
\epsfig{file=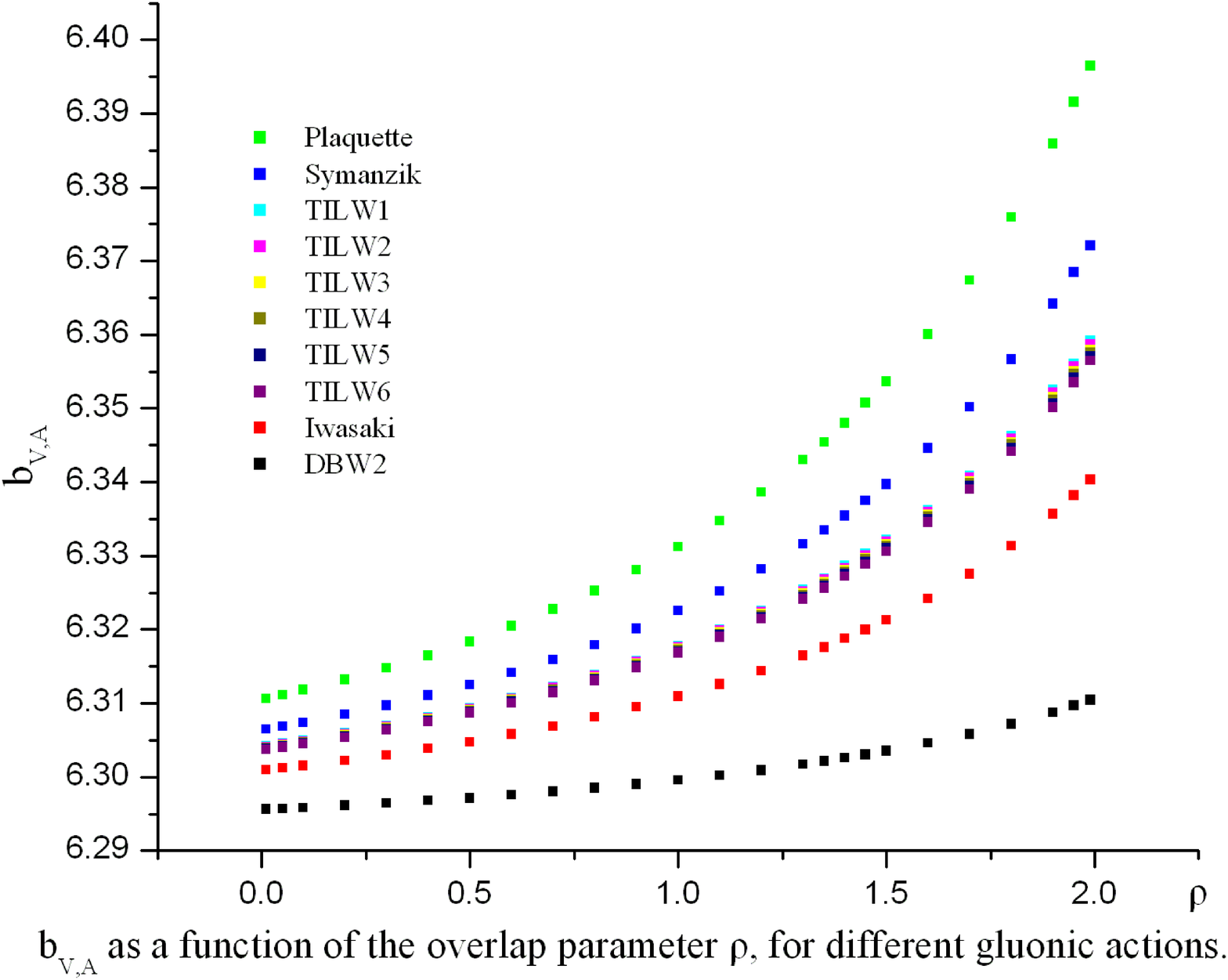, scale=0.15}
\label{vectorpic}
\hfill\\
\hspace{2cm}{\small\bf Figure 3}
\end{center}
\end{minipage}
\hspace{1.3cm}
\begin{minipage}[t]{0.45\linewidth}
\begin{center}
\epsfig{file=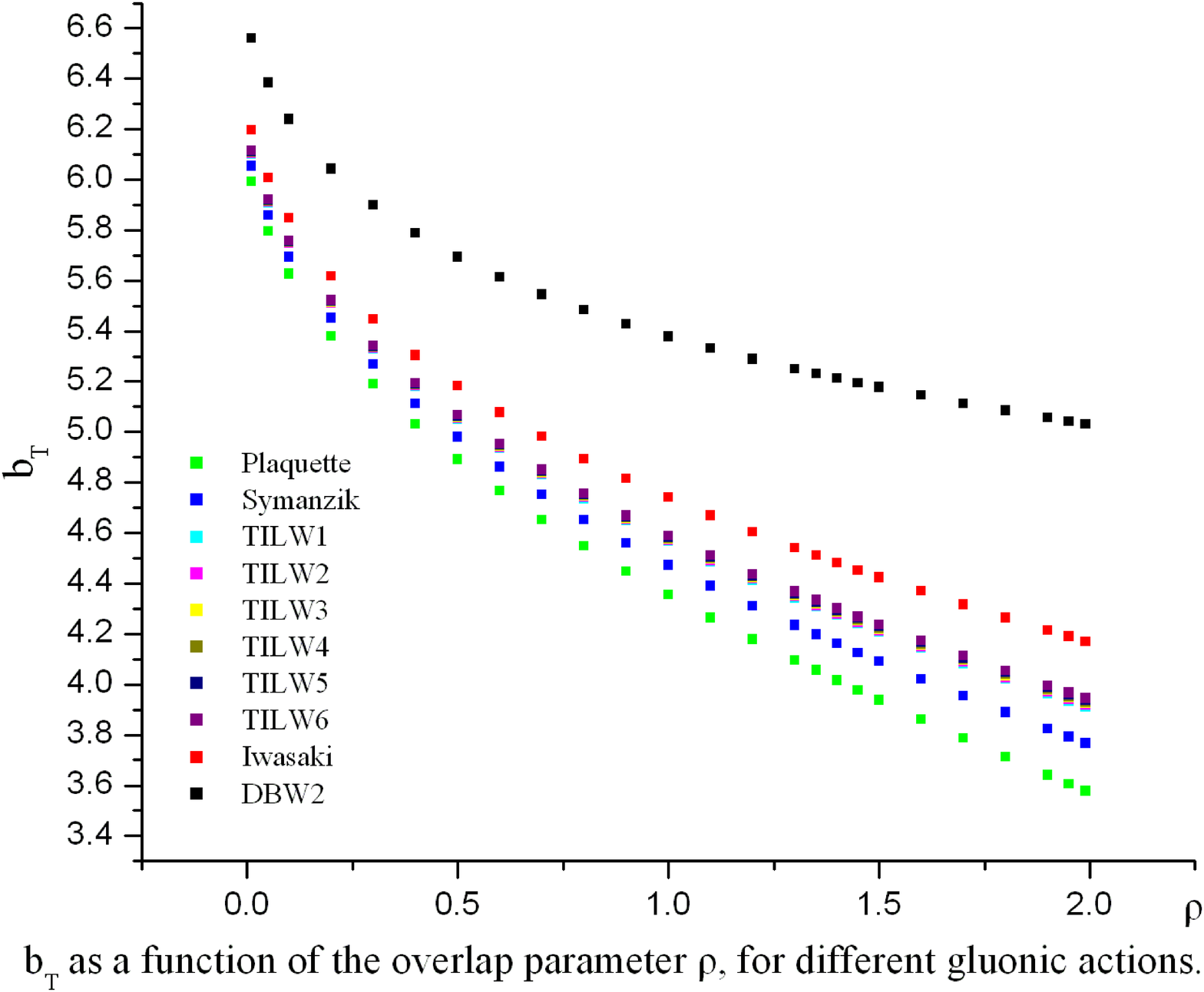, scale=0.15}
\label{tensorpic}
\hfill\\
\hspace{2cm}{\small\bf Figure 4}
\end{center}
\end{minipage}
\end{figure}

Tables of numerical results are presented in our ref \cite{ref2}. In
all cases which can be compared, our results agree with those of
Refs.\cite{ref10,ref3}. There is one exception, regarding the values
of $b_T$  for cases other than the standard plaquette $b_T^{plaq}$ :
In these cases, our results for $b_T-b_T^{plaq}$  have the opposite
sign compared to Ref. \cite{ref3}.\\

\noindent
{\bf First moment of quark distributions}.

 We have calculated the 1-loop renormalization coefficients of the operators $O_{V_2}, O'_{V_2}$  which are the symmetric off-diagonal, and the traceless diagonal parts, respectively, of the operator:
\be
O_{\mu,\nu}=\bar{\psi}\gamma_{\{\mu}D_{\nu\}}\psi
\ee
 The axial counterparts $O_{A_2}, O'_{A_2}$  of the above operators renormalize in the same way. Using the notation of Ref.\cite{ref11}, we find for the amputated Green's function of $O_{\mu,\nu}$ in the Feynman gauge:
\be
\Lambda_{\mu,\nu}(a,p)=\gamma_{\{\mu}p_{\nu\}}+\frac{g^2C_F}{16\pi^2}[(\frac{5}{3}log(a^2p^2)+b)\gamma_{\{\mu}p_{\nu\}}+b'\delta_{\mu\nu}\gamma_{\nu}p_{\nu}-\frac{4}{3}\frac{p_{\mu}p_{\nu}}{p^2}\slashed {p}]
\ee
The quantities $b, b'$  correspond to $(b_1+b_2),\  b_4$  of Ref.\cite{ref3}, respectively. A total of four diagrams (``Tadpole'', ``Vertex'', left and right ``Sails'') contribute in this case.
   The rational coefficients 5/3 and -4/3 in Eq. (3.7) check with
   those of Ref.\cite{ref11}. The values of $b$ and $b'$ are shown in
   Figures 5 and 6; they are in agreement with the quantities
   $b_1+b_2$ and $b_4$, respectively, as defined in Ref.\cite{ref11},
   for the choice of parameters considered in that reference.
\begin{figure}
\hspace{-1.5cm}
\begin{minipage}[t]{0.45\linewidth}
\begin{center}
\epsfig{file=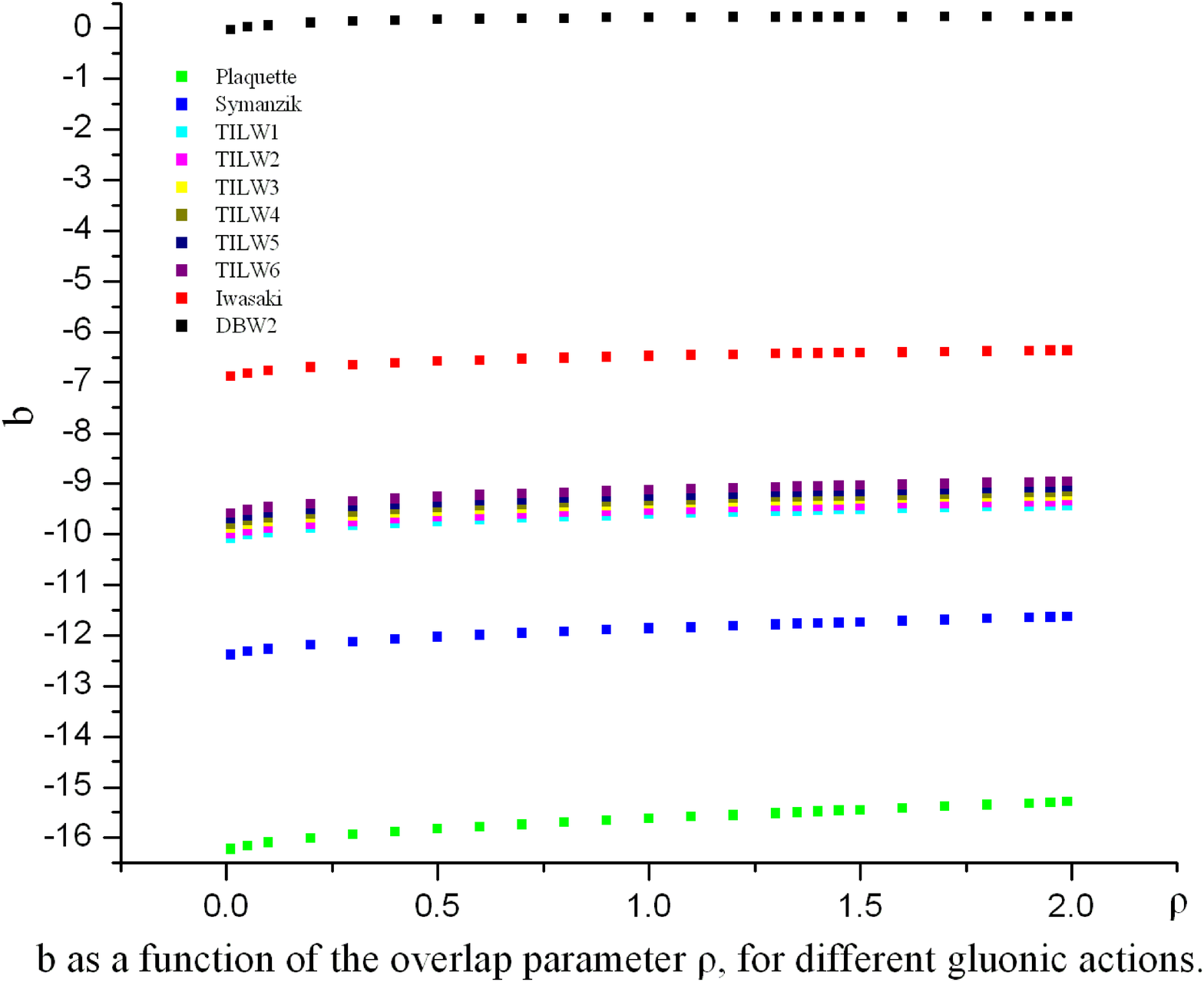, scale=0.15}
\label{b}
\hfill\\
\hspace{2cm}{\small\bf Figure 5}
\end{center}
\end{minipage}
\hspace{1.3cm}
\begin{minipage}[t]{0.45\linewidth}
\begin{center}
\epsfig{file=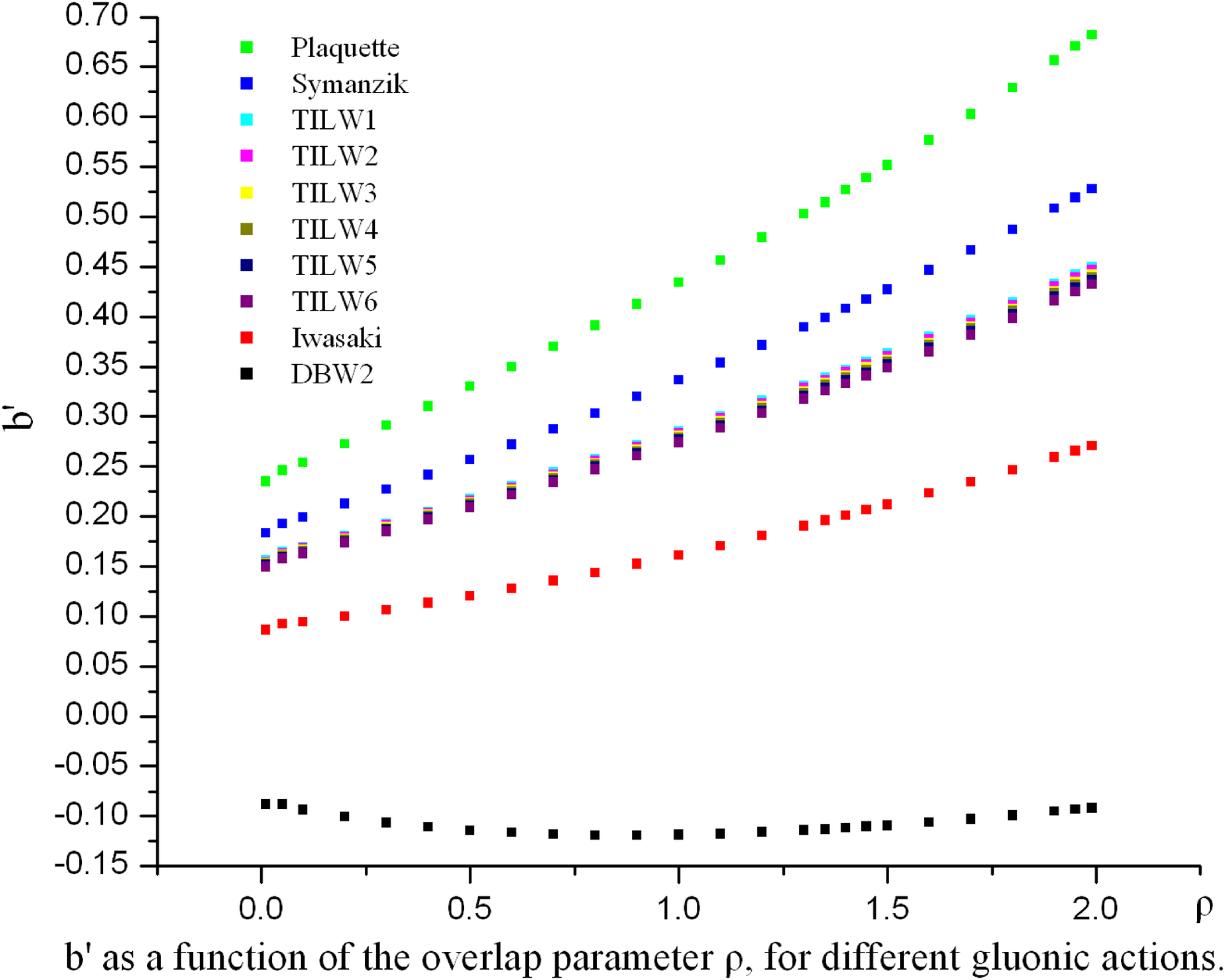, scale=0.15}
\label{bprime}
\hfill\\
\hspace{2cm}{\small\bf Figure 6}
\end{center}
\end{minipage}
\end{figure}  

\eject

\end{document}